# Classification of Instagram fake users using supervised machine learning algorithms

Vertika Singh, Naman Tolasaria, Patel Meet Alpeshkumar, Shreyash Bartwal

SCOPE, VIT University, Chennai Campus, Tamil Nadu, India

**Abstract**

In the contemporary era, online social networks have become integral to social life, revolutionizing the way individuals manage their social connections. While enhancing accessibility and immediacy, these networks have concurrently given rise to challenges, notably the proliferation of fraudulent profiles and online impersonation. This paper proposes an application designed to detect and neutralize such dishonest entities, with a focus on safeguarding companies from potential fraud. The user-centric design of the application ensures accessibility for investigative agencies, particularly the criminal branch, facilitating navigation of complex social media landscapes and integration with existing investigative procedures.

## 1. Introduction

### 1.1 About social media

*Social media, as interactive Web 2.0 Internet-based applications, enable the creation and sharing of ideas, content, and connections. User-generated content forms the essence of social media, with individuals creating profiles to connect with others. While facilitating online social networks, social media also serves various purposes, from learning and self-promotion to memory-keeping and idea development.*

### 1.2 Historical Timeline of social media

*1)* The evolution of social media spans several decades, from early forms of internet communication like email and Bulletin Board Systems (BBS) to the emergence of modern platforms such as Facebook, Twitter, and TikTok. The continuous evolution of social media reflects shifts in user behavior and the introduction of new features and platforms.

## 2. Fake Social-Media (Instagram) Profiles and Their Growing Prevalence

Instagram, ranking third among social media platforms in terms of active users, faces challenges related to fake profiles. These profiles, often associated with identity theft, social engineering, and the spread of harmful content, pose significant security risks to users. Business owners engaging in influencer marketing also encounter issues with overpaying for endorsements due to the prevalence of fake followers. To address these concerns, Instagram employs measures such as algorithms and reporting mechanisms, urging users to be vigilant.

### 2.1 Security Threats Posed by Fake Profiles

Fake profiles not only contribute to identity theft but also serve as conduits for harassment, cyberbullying, fraud, scams, and data harvesting. The extensive use of fraudulent accounts undermines user trust in the platform, affecting the overall user experience and jeopardizing the platform's ability to maintain a safe environment. Instagram utilizes various measures to combat fake profiles, including the examination of user metadata and machine learning algorithms to enhance detection capabilities.

## 3. Proposed Solution: Machine Learning for Profile Verification

To address the challenges posed by fake profiles on social media platforms like Instagram, a potential solution involves examining user metadata and employing machine learning algorithms. These algorithms can analyze posting trends, content engagement, and user behavior to differentiate between genuine and fraudulent profiles. Additionally, the analysis of metadata, including account creation date, IP address history, and device details, provides valuable insights into profile authenticity. Machine learning models, through learning from past data on known fake accounts, offer a robust method for enhancing online social landscape safety and security.

## 4. Literature Review

A variety of methods were employed to classify profiles according to account activity, the quantity of requests that were fulfilled, the quantity of messages that were sent, among other things. System graphs serve as the foundation for the models. Others have attempted to distinguish between cyborgs and robots utilizing certain methodologies. A summary of a few earlier studies is supplied beneath. If specific terms are found in a message, it's deemed unsolicited. This supposition has been employed to identify phony social media accounts. Such Pattern matching algorithms were used to locate phrases on social media. But this criterion falters substantially from the regular creation and use of new terminologies.

Different methods were used to group profiles according to variables including account activity, the number of requests that were answered, the quantity of messages that were delivered, and other characteristics. A framework based on graphs is employed in the models. Others attempted to distinguish between robots and cyborgs using particular techniques. Below is a list of some previous research. Messages can be categorised as spam by using specific terms. This concept has been applied to social media fraud detection profiles. Using pattern matching tools, these phrases were found on social media. However, one major disadvantage of this criterion is the constant generation and application of new nomenclature. Acronyms such as gbu, gn, and LOL are becoming common on Twitter.

As previously mentioned, the concept of a false user can now include human users as well. Many studies, however, only classify phoney users as bot users. The widely used fake project dataset is exclusively made up of bot users; it was obtained through CAPTCHA validation and the market for bot users.Table 1 presents a collection of false users' classifications. The majority of research used Twitter as the platform and employed supervised, feature-based detection techniques. Using several feature sets, supervised machine learning (ML) approaches were used to detect phoney accounts in the fake project dataset. Facebook has more functions connected to media than Twitter does. Identification can be achieved through features like likes (given and received), shares, tags, and comments. There is a report that classified fake accounts on Instagram. Nevertheless, only metadata features were employed, and human judgement was used to determine whether a user was real or fake rather than purchasing fictitious accounts from bot-selling marketplaces. Instagram is becoming more and more popular in influencer marketing, but it doesn't seem to be a popular platform for research.

## 5. Methodology

Data collected from both real and fictitious consumers is the first step in this study. Since media data cannot be obtained from private users, only their metadata could be obtained, all of them were eliminated. Instagram provides the following metadata: username, complete name, bio, link, profile photo, number of posts, following, and followers. These attributes will be extracted following data collection, and a correlation analysis will be performed. Following the configuration of the features, the users will be categorized using machine learning algorithms. In this study four user classes real users and three types of fraudulent users (active, inactive, and spammers) will be discovered. The foundation of these classes is manual behavior observation.

**Block Diagram**

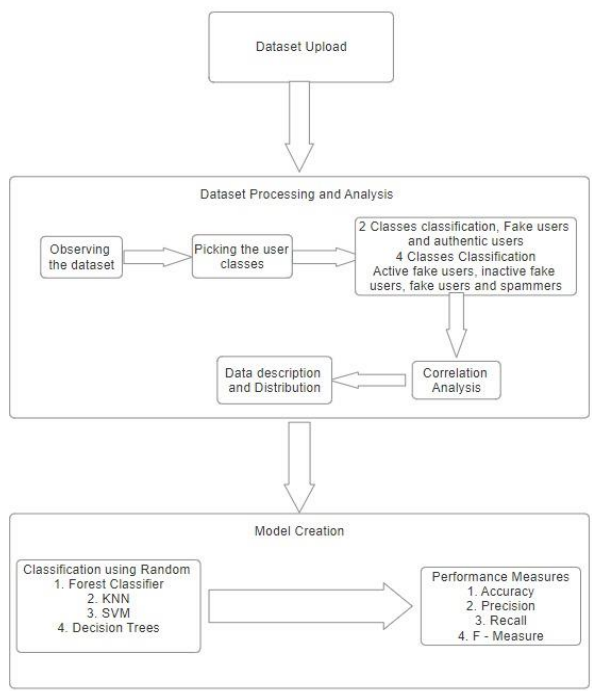

## 6. DATASET

There are five different sources of the features, i.e. metadata, media info, engagement, media tag, media similarity.

1. The Metadata(M) inculcates *pos, flg, flr, bl, pic, lin* -
   - *pos* – the number of posts that the user has posted in total
   - *flg* – number of accounts followed by the user
   - *flr* – number of followers on the user's account
   - *bl* – length of the user's biography
   - *pic* – availability of profile picture on the user's account (Value – 1 if picture is available and 0 if picture is not present)
   - *lin* – availability of any external links (Value – 1 if link is available on the profile and 0 if not)

2. Media Info involves the user's media involving engagement info –
   - *cl* – average length of the captions in the posts or any engagement on the app
   - *cz* – percentage (0.0 – 0.1) of captions that has a length almost equal to zero or non-significant.
   - *ni* – percentage (0.0 – 0.1) of media that doesn't contain any image.

3. ENGAGEMENT INVOLVES THE USER'S ACTIVENESS ON THE APP -

   - *ERL* – IT IS THE ENGAGEMENT RATE WHICH IS DEFINED AS NUMBER OF LIKES DIVIDED BY NUMBER OF MEDIA AND NUMBER OF FOLLOWERS

   - *ERC* – IT IS DEFINED AS THE NUMBER OF COMMENTS DIVIDED BY NUMBER OF MEDIA AND NUMBER OF FOLLOWERS

4. Media tags involves the statistics for tag usage by the user –
   - *lt* – percentage (0.0 - 0.1) of posts tagged with location
   - *hc* – the average count of hashtags used in a post
5. Media similarity –
   - *pr* – average count of promotional keywords (i.e. contest, repost, mention) used in hashtags by the user
   - *fo* – average count of follower hunter keywords (i.e. follow, like, follow for follow) used in hashtags by the user
   - *cs* – average cosine similarity between all pair of two posts a user has

6. interval info –
   - *pi* – average interval between posts (in hours)

## 7.Data Collected

We have used the dataset provided in the paper: Linqing Liu, Yao Lu, Ye Luo, Renxian Zhang, Laurent Itti and Jianwei Lu. "Detecting "Smart" Spammers on Social

Network: A Topic Model Approach." Proceedings of NAACL-HLT. 2016.

**8.Correlation Analysis**

To obtain the correlation table by using the Pearson Correlation values. Upon evaluation we see that there is no strong correlation between any of the variables but two correlation values with the greatest values, bl-lin (0.47) and erc-erl (0.44), are regarded as satisfactory and moderate. The relationship between the length of the biography and the availability of links indicates that users will probably include a link if the biography is lengthy. The comments and the likes correlation demonstrate that there is a linear relationship between the quantity of likes and remarks.
Table present below:

| | pos | flw | flg | bl | pic | lin | cl | cz | ni | erl | erc | lt | hc | pr | fo | cs | pi |
|---|---|---|---|---|---|---|---|---|---|---|---|---|---|---|---|---|---|
| pos | 1.000000 | 0.135049 | 0.061212 | 0.160135 | 0.051938 | 0.169530 | 0.184861 | -0.079381 | 0.077786 | -0.030385 | -0.003858 | 0.029853 | 0.014968 | 0.020762 | -0.010636 | 0.015924 | 0.085687 |
| flw | 0.135049 | 1.000000 | 0.007297 | 0.040520 | 0.010087 | 0.051048 | 0.033444 | -0.022630 | 0.025235 | -0.006179 | -0.008265 | 0.012076 | 0.005407 | 0.015551 | -0.001494 | 0.012064 | 0.013127 |
| flg | 0.061212 | 0.007297 | 1.000000 | 0.009365 | -0.129441 | -0.033437 | -0.063118 | -0.166987 | -0.077394 | -0.023733 | -0.002426 | -0.014543 | 0.004287 | 0.004642 | 0.022982 | 0.022980 | 0.084763 |
| bl | 0.160135 | 0.040520 | 0.009365 | 1.000000 | 0.166268 | 0.471750 | 0.011313 | -0.027234 | 0.143115 | -0.039509 | -0.006439 | 0.021882 | 0.016080 | 0.007192 | 0.017702 | 0.013594 | 0.111549 |
| pic | 0.051938 | 0.010087 | -0.129441 | 0.166268 | 1.000000 | 0.124227 | 0.011568 | -0.062627 | 0.127787 | -0.019804 | -0.002123 | 0.012746 | 0.062906 | 0.022608 | 0.012648 | 0.026452 | 0.084405 |
| lin | 0.169530 | 0.051048 | -0.033437 | 0.471750 | 0.124227 | 1.000000 | 0.030796 | -0.023959 | 0.134877 | -0.004543 | -0.006793 | 0.019967 | 0.009932 | -0.003945 | -0.000953 | 0.009614 | 0.096717 |
| cl | 0.184861 | 0.033444 | -0.063118 | 0.011313 | 0.030796 | 1.000000 | -0.035372 | 0.109931 | 0.003996 | 0.005001 | 0.000801 | 0.018540 | 0.018538 | 0.020778 | 0.055534 | 0.008911 | 0.112708 |
| cz | -0.079381 | -0.022630 | 0.016987 | -0.027234 | 0.062627 | -0.023958 | -0.035372 | 1.000000 | -0.007825 | -0.008550 | 0.000277 | -0.005524 | -0.002216 | 0.006082 | -0.004578 | 0.032024 | 0.063316 |
| ni | 0.077786 | 0.025235 | -0.077394 | 0.143115 | 0.127787 | 0.134877 | 0.109931 | -0.007825 | 1.000000 | -0.002420 | -0.003344 | 0.021913 | 0.004862 | 0.002366 | -0.007850 | 0.024086 | 0.000521 |
| erl | -0.030385 | -0.006179 | -0.023733 | -0.039509 | -0.019804 | -0.004539 | 0.005006 | -0.008550 | -0.002420 | 1.000000 | 0.044350 | -0.004239 | 0.023996 | 0.002453 | 0.034061 | -0.030188 | 0.000753 |
| erc | -0.003858 | -0.008265 | -0.002426 | -0.006439 | -0.002123 | -0.006739 | 0.000601 | 0.000277 | -0.003344 | 0.044350 | 1.000000 | 0.043567 | 0.014676 | 0.045538 | 0.024061 | 0.031888 | 0.000753 |
| lt | 0.029853 | 0.012076 | -0.014543 | 0.021882 | 0.012746 | 0.019967 | 0.008401 | -0.005524 | 0.021913 | -0.004239 | 0.043567 | 1.000000 | 0.014675 | 0.045538 | 0.024061 | 0.031888 | 0.000753 |
| hc | 0.014968 | 0.005407 | 0.004642 | 0.016080 | 0.062906 | 0.009394 | 0.018538 | -0.002216 | 0.004806 | 0.023996 | 0.014675 | | | | | | |
| pr | 0.020762 | 0.015551 | 0.004642 | 0.007192 | 0.022608 | -0.003945 | 0.020778 | 0.006082 | 0.023015 | 0.002453 | 0.045538 | | | | | | |
| fo | -0.010636 | -0.001494 | 0.022982 | 0.017702 | 0.012648 | -0.000953 | 0.055534 | -0.004578 | -0.007850 | 0.034061 | | | | | | | |
| cs | 0.015924 | 0.012064 | 0.022980 | 0.013594 | 0.026452 | 0.009614 | 0.008981 | 0.032024 | 0.024068 | -0.030188 | | | | | | | |
| pi | 0.085687 | 0.013127 | 0.084763 | 0.111549 | 0.084405 | 0.096717 | 0.112708 | 0.063316 | 0.000521 | 0.000753 | | | | | | | |

## Correlation Matrix (continued)

| | erc | lt | hc | pr | fo | cs | pi |
|---|---|---|---|---|---|---|---|
| erc | -0.038688 | 0.00082 65 | -0.00242 67 | -0.00604 38 | -0.00612 31 | -0.00673 79 | -0.00506 17 |
| lt | 0.029853 | 0.01207 6 | 0.01182 54 37 | -0.21746 1 | 0.12748 | 0.19687 | 0.08401 0 |
| hc | 0.014968 | 0.00540 7 | -0.00420 87 | 0.16080 8 | 0.06290 6 | 0.09382 0 | 0.18538 3 |
| pr | 0.020762 | -0.00155 1 | -0.04664 2 | -0.02719 2 | -0.02608 | -0.03945 3 | 0.20781 8 |
| fo | -0.010636 | -0.00149 4 | 0.02298 2 | 0.01770 2 | 0.00124 8 | -0.00953 5 | 0.05534 |
| cs | -0.015924 | -0.01206 4 | 0.22908 0 | -0.13591 4 | -0.26452 6 | -0.09614 7 | -0.08981 1 |
| pi | -0.085687 | -0.00131 27 | -0.00847 63 | -0.01114 49 | 0.08405 17 | -0.09067 08 | 0.01127 |

### Data description and distribution

| | pos | flw | flg | bl | pic | linc | cl | cz | ni | erl | erc | lt | hc | pr | fo | cs | pi |
|---|---|---|---|---|---|---|---|---|---|---|---|---|---|---|---|---|---|
| count | 6532 6.000 000 | 6.5326 e+04 | 6532 6.000 000 | 6532 6.000 000 | 6532 6.000 000 | 6532 6.000 000 | 6532 6.000 000 | 6532 6.000 000 | 6532 6.000 000 | 6532 6.000 000 | 6532 6.000 000 | 6532 6.000 000 | 6532 6.000 000 | 65326.000000 | 6.5326 e+04 | 6532 6.000 000 | 6532 6.000 000 |
| mean | 176.57 1227 | 1.1830 77e+03 | 2310. 5192 1 | 57.49 7065 | 0.951 725 | 0.282 819 | 136.6 5203 44 | 0.250 012 | 0.193 229 | 19.14 6641 | 1.1394 21 | 0.1308 877 | 0.2050 7796 | 327.45 | 6.0552 837 | 0.29 9098 | 496.4 7519 3 |
| std | 723.47 0655 | 2.1708 02e+04 | 2592. 0961 04 | 64.12 9264 0 | 0.214 4264 | 0.449 9940 | 215.7 1444 86 | 0.333 7891 | 0.252 5239 | 121.0 4753 0 | 5.8106 627 | 0.3030 3362 | 1.1569 21 | 0.20987 | 0.5190 5558 | 0.34 9604 | 944. 9053 89 |
| min | 0.000 000 | 0.000 00e+00 | 0.000 00 | 0.000 00 | 0.000 00 | -1.000 00 | 0.000 00 | 0.000 00 | 0.000 00 | 0.000 00 | 0.000 00 | 0.000 00 | 0.000 00 | 0.000000 | 0.000 00 | 0.00 000 | 0.00 0000 |
| 25% | 6.000 000 | 1.2300 e+02 | 394. 0000 | 0.000 00 | 1.000 00 | 0.000 00 | 8.000 00 | 0.000 00 | 0.000 00 | 2.7308 | 0.0000 | 0.0000 | 0.0000 | 0.000000 | 0.000 00 | 0.03 3327 | 24.5 7145 8 |
| 50% | 30.00 000 | 3.3800 e+02 | 995. 0000 | 32.0 000 | 1.000 00 | 0.000 00 | 46.0 0555 6 | 0.050 00 | 0.055 900 | 9.4500 | 0.4400 | 0.0000 | 0.0077 00 | 0.000000 | 0.000 00 | 0.13 6915 | 183. 2279 43 |
| 75% | 124.0 0000 | 8.1700 e+02 | 3600 .000 00 | 110. 0000 | 1.000 00 | 1.000 00 | 170. 4444 44 | 0.443 300 | 0.333 080 | 18.6 0433 0 | 1.0433 30 | 0.0061 10 | 0.0000 00 | 0.000000 | 0.000 00 | 0.45 6342 | 580. 7719 27 |
| max | 7620 0.000 000 | 3.9000 e+06 | 8800 0.000 000 | 555. 0000 | 1.000 00 | 1.000 00 | 3644 .000 000 | 1.000 00 | 1.000 00 | 2665 0.902 7 | 1009. 0000 | 1.0000 00 | 30.0 0000 0 | 20.000000 | 58.0 0000 | 1.00 000 | 2678 6.134 766 |

The Correlation Matrix shows the measure of dependency of variables. As seen in the table no strong correlation can be seen.

Correlation between lin(external link) and bl(bio length) is moderately high(0.471750) this tells us that the users with long bio will most likely put a link in their profile. Erc and erl also show moderate correlation(0.443567) this means that the number of likes is linearly related to number of comments. other variables with high correlation which is on the weaker side include:

Cz and cl with a correlation of -0.353722, the purpose of the cz is to enhance the cl as they are inversely dependent. Typically, fake users will upload media with almost no caption> in the dataset the percentage of fake users with almost zero caption is high.

Correlation between bl and cl is 0.350113, users with lengthy captions appear to have a higher association between bl (biography length) and cl (caption length) as they tend to have longer biographies help in identifying authentic accounts.

Fo and hc also have good correlation as fo is a subset of hc.

**Statistics of features available in the two datasets-**

Metadata (pos, bl, image, lin): The quantity of posts made by fake users is nearly equal to that of real users. The spammers, however, have a notably larger number of posts. The longest biography text is seen among authentic users, and many of them include a link. Of inactive users, 76% of them just have a profile while other classes all nearly have profile pictures.

Follow info (flw, flg): Authentic users have the highest followers count, but lowest following count. In contrast, fake users have a lower followers count, but a higher following count if compared to the authentic users. This indicates that fake users like to follow others to increase their presence.

Engagement (erl, erc): When compared to real users, fake users will get more likes. On the other hand, genuine users get more comments. This suggests that getting comments is more difficult, thus helping in finding the real users. Fake users typically follow other fake users in terms of likes, in order for them to get automatic likes.

Media information (cl, cz, ni): Real users have more detailed captions and less zero captions than fraudulent users.

Media tags (lt, hc): When compared to fraudulent users, real users utilize location tags and hashtags more frequently. To draw users, spammers utilize more hashtags.

- Media similarity (pr, fo, cs, pi): A lower cs (cosine similarity) value is associated with authentic users. Thus, the majority of their posts are distinct from their earlier postings, in contrast to phony users. Those who spam have the highest FO(follow keywords) and PR(promotional keywords).

| class | pos | flw | flg | bl | pic | lin | cl | cz | ni | erl | erc | lt | hc | pr | fo | cs | pi |
|---|---|---|---|---|---|---|---|---|---|---|---|---|---|---|---|---|---|
| a | 186.02 97 83 | 82 9.5 37 41 5 | 31 42. 12 78 41 | 51 .1 24 19 1 | 0.98 98 79 | 0.14 94 11 | 117. 73 93 40 | 0.29 82 02 7 | 0.22 72 96 | 9.85 72 4 | 0.43 06 34 | 0.17 08 69 9 | 0.34 87 03 | 0.01 42 97 9 | 0.04 28 57 | 0.22 22 85 25 | 398.96 48 86 |
| i | 1.80 80 39 | 215.7 14 28 6 | 38 16. 52 20 40 | 12.4 75 87 4 | 0.76 54 7 5 | 0.03 06 1 9 | 11.73 05 75 4 | 0.34 41 56 8 | 0.05 68 39 | 38.2 05 26 5 | 1.85 96 1 8 | 0.04 70 2 3 | 0.04 71 05 8 | 0.00 10 7 3 | 0.00 04 5 5 | 0.64 24 74 5 | 315.99 67 23 |
| r | 123. 94 52 16 | 11 93. 70 80 74 | 12 07. 97 82 | 61.8 82 57 8 | 0.98 62 08 | 0.22 12 4 2 | 12.67 85 75 | 0.25 04 8 6 | 0.28 45 61 2 | 20.5 70 57 5 | 1.53 34 7 5 | 0.34 47 61 2 | 0.07 46 16 9 | 0.04 68 9 8 | 0.00 38 65 4 | 0.16 65 54 3 | 731.10 13 68 |
| s | 297. 27 90 61 | 10 81. 28 57 84 | 30 52. 56 94 24 | 59.3 92 77 0 | 0.98 27 5 | 0.16 66 4 | 23 6.93 91 99 | 0.31 67 8 5 | 0.16 28 24 0 | 14.3 28 22 | 1.72 13 30 | 0.17 76 28 | 0.16 05 89 | 0.01 76 7 6 | 0.27 27 74 8 | 0.30 76 96 9 6 | 327.96 84 66 |

## 9. Machine Learning Models used

We have used five machine learning algorithms for the classification in our research. We have basically done two types of classification, the first one being a 2-class classification for detecting the real and fake users and the second one being a 4-class classification for segregating the users into 4 classes namely active fake users, inactive fake users, spammers and authentic users.
The algorithms used by us are Random Forest Classifier, KNN, SVM and Decision Trees.

## 10. Model Analysis

| s.no. | Algo | 2 class classification | | | | 4 class classification | | | |
|---|---|---|---|---|---|---|---|---|---|
| | | acc | prec | recall | F1 | acc | Prec | recall | F1 |
| 1 | Random forest | 89.63 | 91 | 90 | 90 | 89.15 | 89 | 89 | 89 |
| 2 | Knn | 74.05 | 74 | 74 | 74 | 55.10 | 56 | 55 | 55 |
| 3 | Svm(polynomial) | 63.95 | - | - | 60.09 | 36.18 | - | - | 60.09 |
| | Svm(rbf) | 53.83 | - | - | 41.59 | 29.64 | - | - | 41.59 |
| 4 | Decision trees | 87.24 | 89 | 87 | 87 | 72.22 | 72 | 72 | 71 |

Random forest has given better results in every aspect so we have used that in our project. Random Forest even performs well in the 4-class classification while other algorithms find it difficult. The top four predictors in the 4-class categorization are pos, flw, bl, and flg. The top five predictors in the two-class categorization are pos, flw, lin, flg, and bl. All of these metadata values are easily obtainable, even for private users hance making our model successful even without all the values.
We achieve high accuracy in 2 class classification unlike in the 4 class classification. The 4-class classification's lower accuracy result can be attributed to the significant dependence on media data for the distinguishing of fake user types.

## 11. Results with Discussion

Hence upon training and testing various ML models we can conclude that using Random Forest model we can achieve the highest accuracy and is the effective model that can be used to distinguish between authentic and fake users.
The reasons why Random Forest model stood out are as follows:
1. Ensemble Method: It is a member of the ensemble method family, which builds on the strengths and accuracy of individual models to produce a more robust model. By combining the predictions from several decision trees, Random Forest lowers the chance of overfitting and boosts accuracy.
2. Reduction of Overfitting: The Random Forest method uses bootstrapping and feature randomness to train each decision tree on a distinct subset of data and features. Because of the diversity and randomness among the trees, overfitting is less common, and the model performs better on unknown data.
3. Managing Non-linear Relationships: Non-linear relationships between features and the target variable can be captured by Random Forests. The system is capable of managing intricate relationships found in the data by utilizing multiple trees and taking into account distinct feature subsets for every tree.

**12.Conculsion**

At the end of the research, we can say that to ensure authenticity of users on Instagram one should take text and image analysis into consideration, the texts users use in their captions and comments for posts might not be relevant at times which may help in fake user identification. Image analysis and interval between posts analysis is also very useful in identifying spammers. Hence to make Instagram a better platform the above steps can be taken into consideration.